\documentclass{svmult}
\pdfoutput=1

\usepackage{mathptmx}       
\usepackage{helvet}         
\usepackage{courier}        
\usepackage{type1cm}        
\usepackage{makeidx}         
\usepackage{graphicx}        
\usepackage{multicol}        
\usepackage[bottom]{footmisc}

\usepackage{amsmath}

\makeindex             

\begin{document}

\title*{Individual Microscopic Results Of Bottleneck Experiments}
\author{Marek Buk\' a\v cek and Pavel Hrab\' ak and Milan Krb\' alek}
\institute{Faculty of Nuclear Sciences and Physical Engineering, Czech Technical University in Prague, Trojanova 13, Prague 2, 120 00, Czech Republic.
}

%
%
\maketitle

\abstract{This contribution provides microscopic experimental study of pedestrian motion in front of the bottleneck, explains the high variance of individual travel time by the statistical analysis of trajectories. The analysis shows that this heterogeneity increases with increasing occupancy. Some participants were able to reach lower travel time due more efficient path selection and more aggressive behavior within the crowd. Based on this observations, linear model predicting travel time with respect to the aggressiveness of pedestrian is proposed.}

\section{Experiment}

Various experiments have been conducted in order to verify crowd behavior models and to enable fundamental research of pedestrians phenomena \cite{Schadschneider2010,Steffen2010,
Seyfried2010a,Seyf2014}.

Advanced processing of video records provides microscopic analysis of individual behavior \cite{TGF13,Duivf2014}. During critical situation, the individual behavior plays important role -- less aggressive pedestrians spend more time in monitored area, which may cause unexpected complications. 

\begin{figure}[h!]
	\centering
	\includegraphics[height=3.7cm]{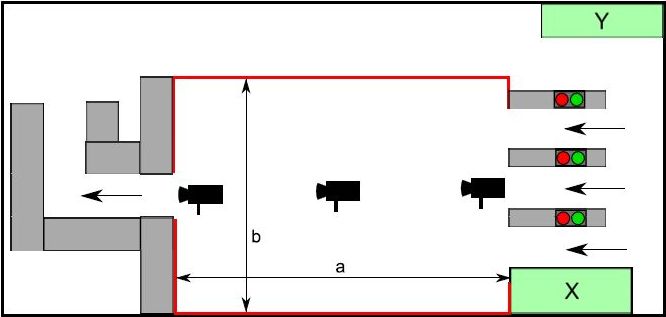}
	\phantom{x}
	\caption{Left: schema of organized experiment. The distance from the entrance to the exit a = 7.2 m (considered for measuring of travel time), room width 4.5 m and exit width 0.6 m. Right: snapshot from exit camera.}
	\label{fig:room} 
\end{figure}

This article is based on egress experiment organized in Czech Technical University. Group of 75 students passes through artificial room (Fig. \ref{fig:room}), instructed to leave the area as fast as possible, to avoid running and pushing each other. The results supported phase transition study mentioned in \cite{PED14,ACRI14}.

Tree entrances were controlled by traffic lights to get demanded traffic mode inside the experimental area, see Table \ref{tab:makro}. To simulate random inflow conditions, green light was alternated by $k\cdot \Delta h$ seconds of red light, where $k$ was generated from geometric distribution and $\Delta h = 0.6$ s was the time step. Each round started with empty room. 

Unique codes on the hats of participant enabled to detect and identify trajectory of each participant. From this information, travel time (covering the period from entrance to exit) and Voronoi density were extracted for further investigation.

\begin{table}
	\caption{Summary of performed rounds. $J_{\rm in}$ and $J_{\rm out}$ refer to flow measured flow at the entrance, resp. at the exit, $\overline{TT}$ is mean travel time in given round. $N(150)$ specify the number of pedestrians in the room 150 s after initialization and \# paths denotes the number of passings in given round.}
	\label{tab:makro}
\begin{tabular}{rcccccl}
	round	& $J_{\rm in}$ [ped/s] & $J_{\rm out}$ [ped/s] & $\overline{TT}$ [s]& $N(150)$ [ped] & \# paths & Observation\\
	\hline\hline	
	\#~~2 &0.99 &0.99 &\hspace{1.1ex}5.67 &\hspace{1.1ex}3 &158 &free flow\\
	\#~~5 &1.22 &1.20 &\hspace{1.1ex}6.73 &\hspace{1.1ex}7 &274 &free flow\\
	\#~~4 &1.37 &1.30 &16.59 & 24  &294 &stable cluster\\
	\#~~3 &1.43 &1.33 &14.39 & 22  &260 &stable cluster\\
	\#~~6 &1.39 &1.31 &20.40 & 33  &270 &stable cluster\\
	\#~~7 &1.55 &1.37 &25.78 & 45  &260 &transition\\
	\#~11 &1.61 &1.38 &21.65 & 41  &141 &transition\\
	\#~~9 &1.78 &1.37 &24.06 & 47* &148 &congestion\\
	\#~~8 &1.79 &1.38 &25.03 & 46* &144 &congestion\\
	\#~10 &1.78 &1.37 &23.33 & 44* &214 &congestion\\
	\hline
\end{tabular}
\end{table}

\section{Travel Time Analysis}
As mentioned in \cite{PED14}, recorded travel time significantly depends on the occupancy $N(t)$, defined as number of pedestrians inside the room. Travel time increases linearly with occupancy and the variance is increasing as well, see Figure \ref{fig:TT_paths} below. This phenomenon will be described by pedestrian's individual characteristics. 

The term trajectory is understood to be the set of time-space coordinates assigned to one participant during one of his passings
$$
\vec{x}_i = {(x_i(t),y_i(t),t)},
$$
where $x_i(t)$ and $y_i(t)$ are coordinates of paths $i$ in time $t$. Here we note that the path identifier $i$ does not refer to any specific participant, but to the recorded trajectory.

Travel time of trajectory $i$ is defined as the time spent in the room, i.e.
$$
TT(i)=T_\mathrm{out}(i)-T_\mathrm{in}(i),	
$$
where the $T_\mathrm{in}$ and $T_\mathrm{out}$ is measured just behind entrances, resp in front of the exit, see Figure \ref{fig:room}.

The occupancy in the room $N(t)$ was derived from paths, this quantity was used to determine mean occupancy $\overline{N}{(i)}$ for each path $i$: as
$$
\overline{N}(i)=\frac{1}{TT(i)}\int_{T_\mathrm{in}{(i)}}^{T_\mathrm{out}{(i)}}N(t)\mathrm{d}t
$$ 
     	
\begin{figure}[h!]
	\centering
	\includegraphics[height=6cm]{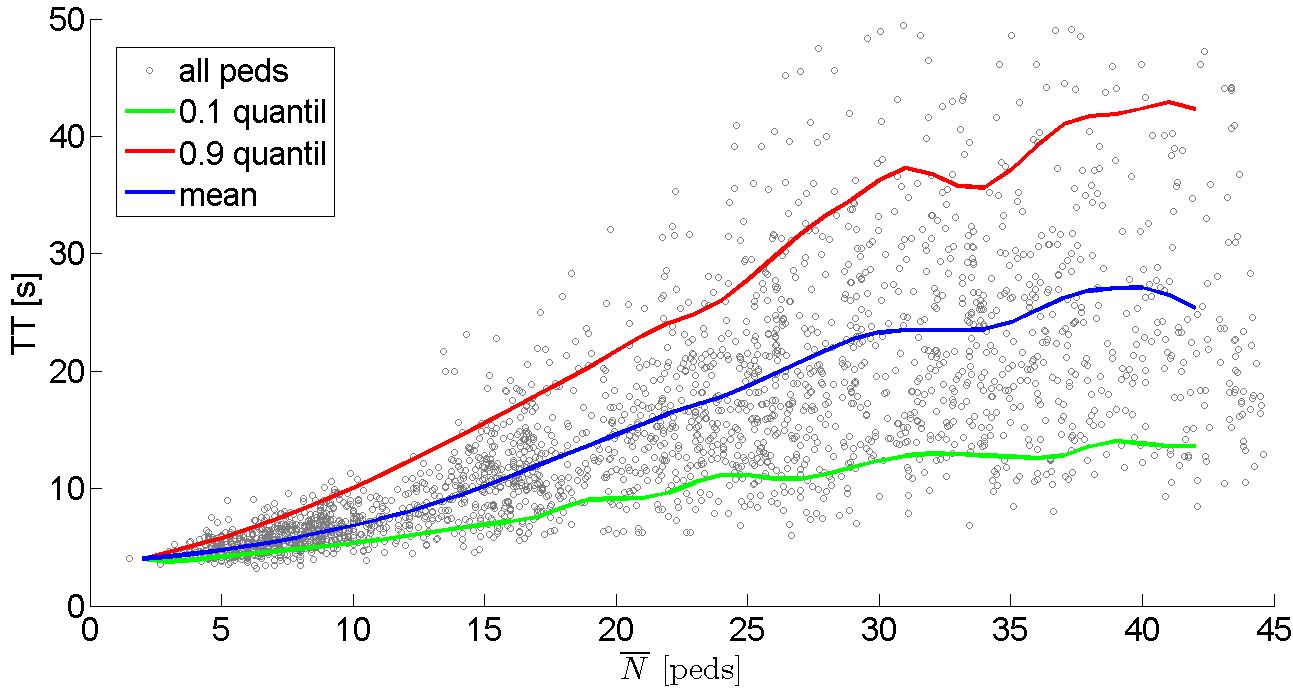}  
	\caption{Travel time -- occupancy dependency. Each point represents one passing, lines visualize mean and quantiles evaluated for given occupancy.}
	\label{fig:TT_paths}
\end{figure}

\subsection{Relative travel time}
To compare the travel times measured under different conditions (stable state was not reached for whole experiment), scaling based on mean occupancy was introduced. For each occupancy bin $(N-1,N\rangle$ the mean travel time $TT_N$ was defined as
$$
TT_N = \operatorname*{mean}\limits_{i} \left\{TT(i) \mid \overline{N}(i)\in (N-1,N]\right\}\,.
$$
Then the relative travel time for each path may be evaluated as
$$
TT_R(i) = \frac{TT(i)}{TT_N}\,,
$$
that enable to compare Travel time of paths reached under different conditions.

\begin{figure}[h!]
	\centering
	\includegraphics[height=3cm]{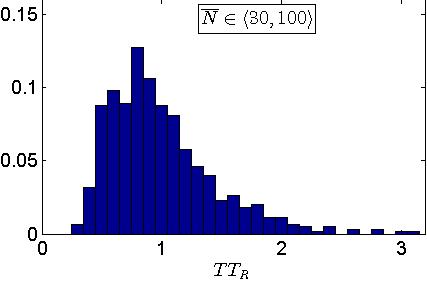}  
	\phantom{x}
	\includegraphics[height=3cm]{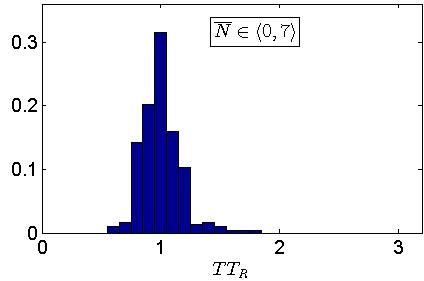} 
	\caption{Histograms of relative travel time, data filtered for high density (left) and low density (right) areas.}
	\label{fig:RelTT_paths}
\end{figure}

To underline the increase of variance in travel time, two histograms of relative travel time are plotted in Figure \ref{fig:RelTT_paths}. As you can see, the travel time of dominant majority of passings in free flow did not deviate more than 20\% from the mean value. On the other hand, the travel time in congested mode covered area from 50\% to 200\% of $TT_N$.

\subsection{Paths and Paths Density}
Paths selection is one of the features, which affect reached travel time. Following study summarizes the space usage described by paths density. The workload $W$ of space element $A_{j}$ is defined as
$$
W(A_{j}) = |\left\lbrace i: \exists t \in \vec{x}_i: (x(t),y(t) \in A_{j} \right\rbrace|	\,,
$$
where $A_{j}$ is an area defined by rectangular grid, $0.2$ m $\times$ 0.2 m.
 
\begin{figure}
\newlength{\w}
\setlength{\w}{2.8cm}
\newlength{\h}
\setlength{\h}{2.3cm}
\begin{tabular}{cccc}
	fast in metastable & slow in metastable & fast in crowd & slow in crowd \\
	\includegraphics[height=\h, width=\w]{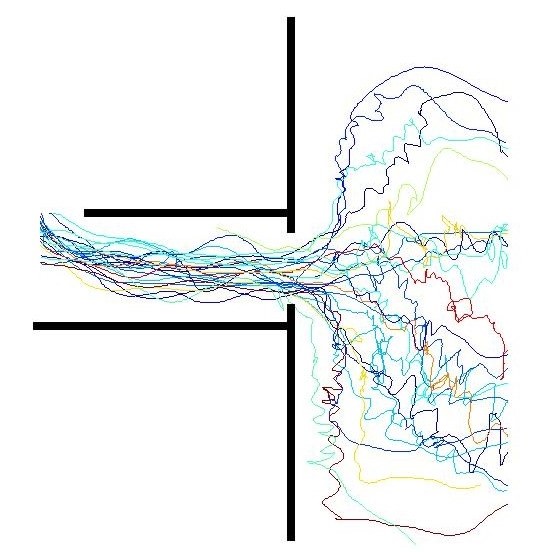} &
	\includegraphics[height=\h, width=\w]{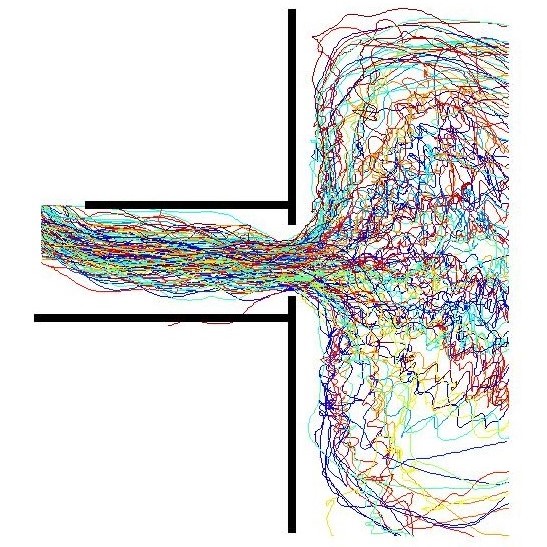} &
	\includegraphics[height=\h, width=\w]{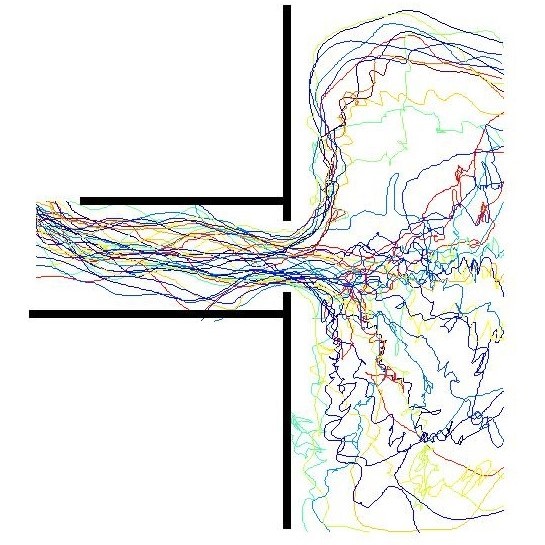} &	
	\includegraphics[height=\h, width=\w]{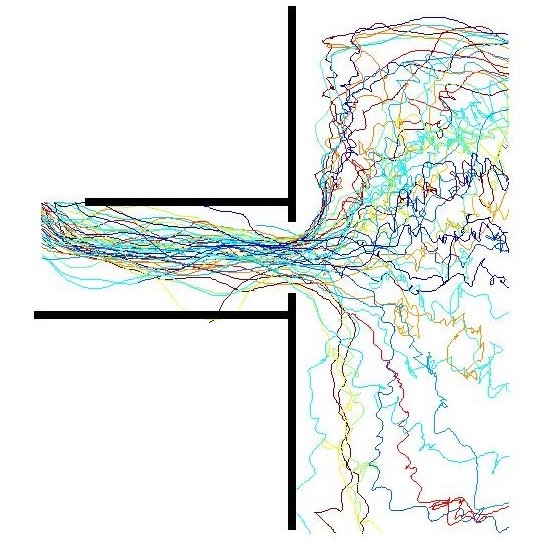} \\
	\includegraphics[height=\h, width=\w]{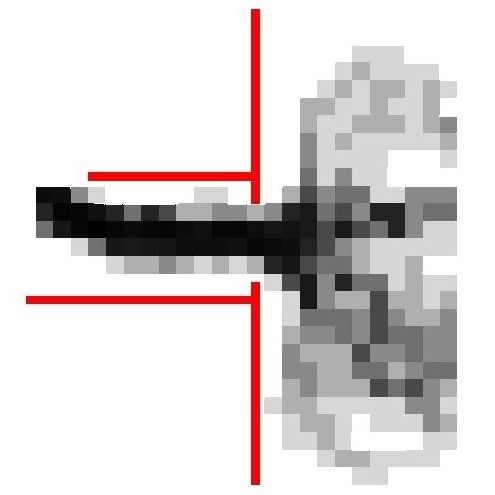} &
	\includegraphics[height=\h, width=\w]{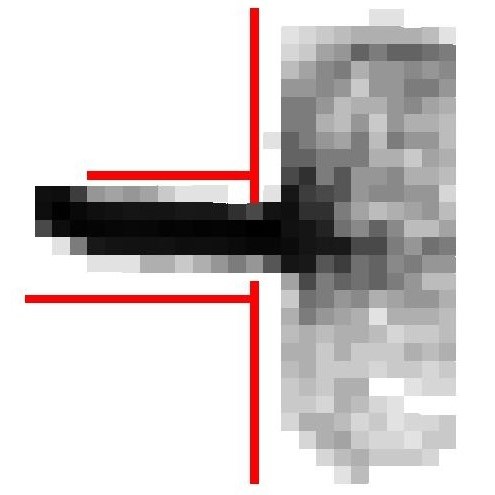} &
	\includegraphics[height=\h, width=\w]{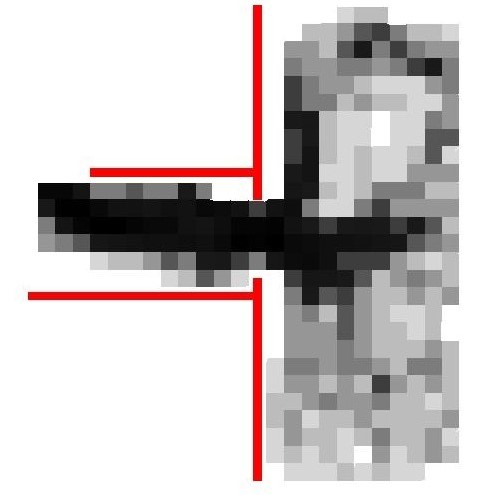} &
	\includegraphics[height=\h, width=\w]{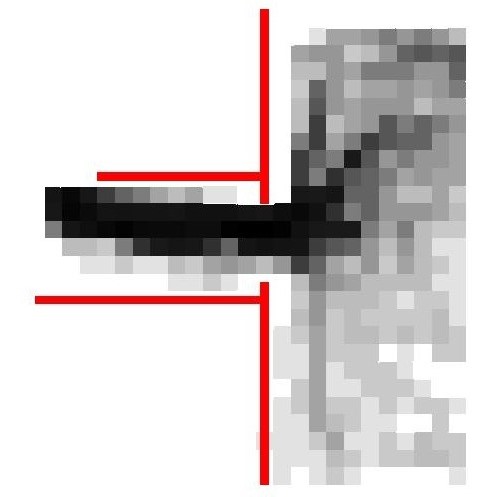} \\[2mm]
	 $\overline{N}\in[25,35]$&$\overline{N}\in[25,35]$&$\overline{N}\in[35,50]$&$\overline{N}\in[35,50]$ \\[2mm]
	 $TT\leq10$ & $TT\geq21$ & $TT\leq15$ & $TT\geq35$ 
\end{tabular}
	\caption{Paths (first row) and path density (second row) evaluated for two different density areas: metastable cluster and congested crowd. For both traffic modes, trajectories of slow and fast pedestrians were compared. Here we note that the density was evaluated on a grid 0.2 m $\times$ 0.2 m, each trajectory contributed to any segment maximally ones. The darker color, the  higher workload.}
	\label{fig:group_paths}
\end{figure}

Both trajectory and workload were evaluated in area in front of the bottleneck, the visualization is provided in Figure \ref{fig:group_paths}. In the following the terms slow and fast trajectories refer to set of 20\% higher, resp lower  travel time. These trajectories are compared under meta-stable and congested phase of the system, several conclusions were drawn:
\begin{itemize}
	\item the direct path to the exit was used more by fast trajectories,
	\item the area at the wall was used by fast paths,
	\item slow paths observed mainly along the angle 45$\deg$ to the exit,
	\item observed asymmetry - fast on left, slow on right.
\end{itemize}

\subsection{Individual approach}

More detail study may be provided adding the pedestrian's identification. Here we will use the Greek letters to denote identified participants.

While some participants reached similar travel time in free flow and congested mode, the others were not able to pass through the dense crowd and spent incomparable more time in the room.

To compare participants, individual relative travel time was defined as
$$
\overline{TT}_{\alpha} = \operatorname*{mean}\limits_{i} \left\{TT_{R}(i) \mid i \in I_{\alpha} \right\},
$$
where $I_{\alpha}$ is the set of paths assigned to pedestrian $\alpha$. 

\begin{figure}[h!]
\begin{center}
	\includegraphics[height=3cm]{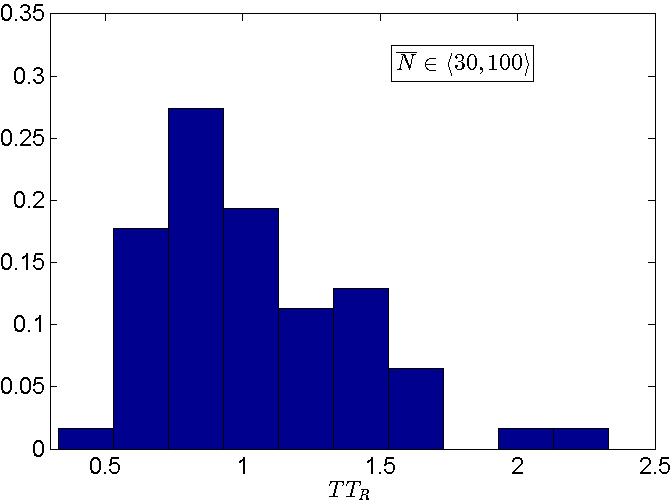}
	\phantom{x}
	\includegraphics[height=3cm]{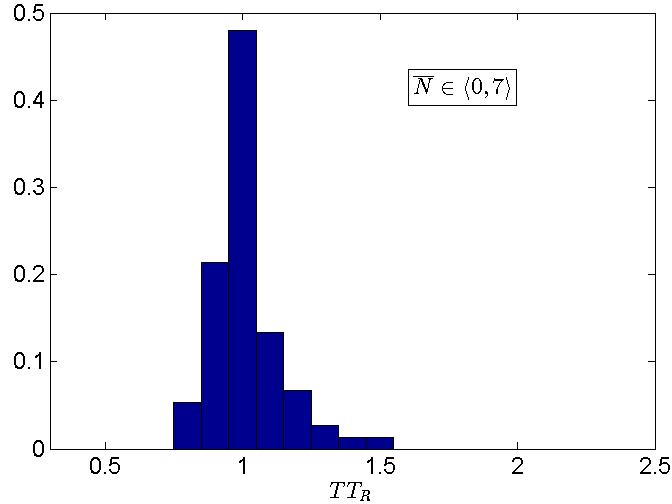}
	\caption{Histograms of individual relative travel time. Data from free flow mode (right) are compared to high occupancy periods (left).}
	\label{fig:hist_individual}
\end{center}
\end{figure}

The histograms of $\overline{TT}_{\alpha}$ for free flow and congested state are visualized in Figure~ \ref{fig:hist_individual}. As you can see, the heterogeneity among participants corresponds well to the variance of relative travel time introduced in Fig. \ref{fig:RelTT_paths}.

\begin{figure}[h!]
	\begin{center}
		\includegraphics[height=0.25\textheight]{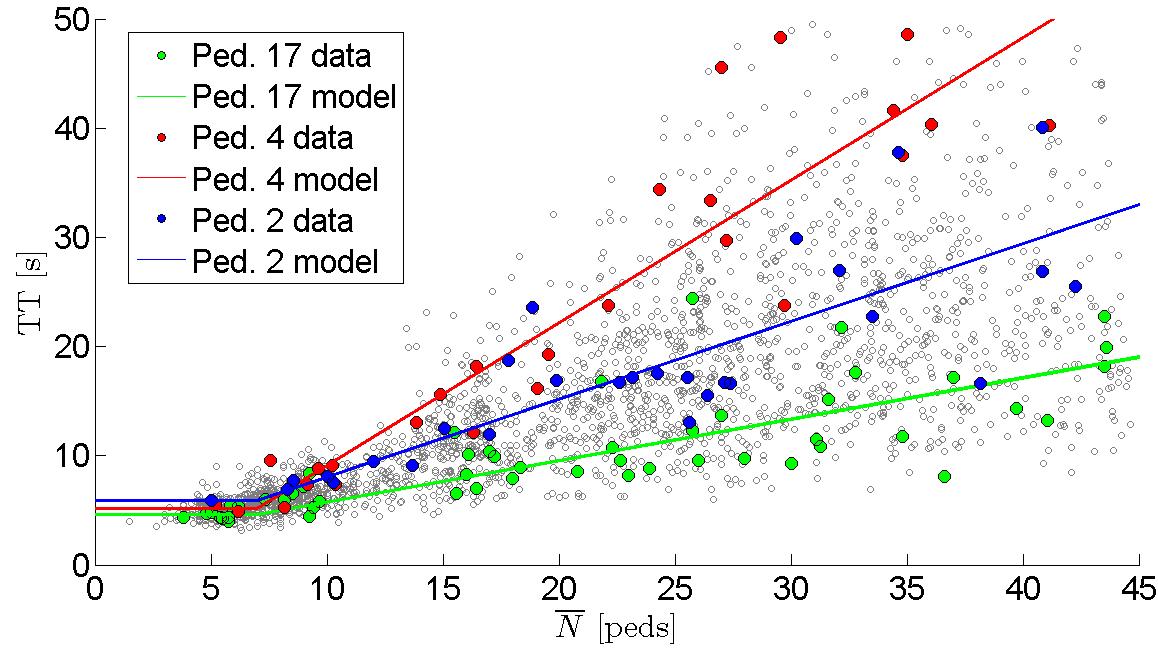}
	\end{center}
	\caption{Individual travel time with respect to the mean occupancy. Data for three pedestrians with different strategies were highlighted and piece-wise linear model is illustrated.}
	\label{fig:TT_individual}
\end{figure}

Thanks to participant identification it is possible to highlight records corresponding to given pedestrian in the travel time -- occupancy diagram (Figure \ref{fig:TT_individual}). The observations show the same heterogeneity measured by individual relative travel time. While some participant were not affected much by crowd, some were not able to reach the exit through crowded area.

To compare the ability to push through the crowd, we define the pedestrian attribute aggressiveness as the slope of piece-wise linear model  
$$	
TT(i_{\alpha})=\frac{S}{v_0} + \mathbf{1}_{\{\overline{N}(i_{\alpha})>7\}}(\overline{N}(i_{\alpha})-7) \cdot \mathrm{slope}(i_{\alpha}) + \mathrm{noise}.
$$			
The factor ($\overline{N} > 7$) specifies the mode, where is one pedestrian affected by motion of others. Until the $\overline{N} < 7$, the free phase is observed and therefore the interactions may be neglected.

\begin{figure}[h!]
	\begin{center}
		\includegraphics[height=0.15\textheight]{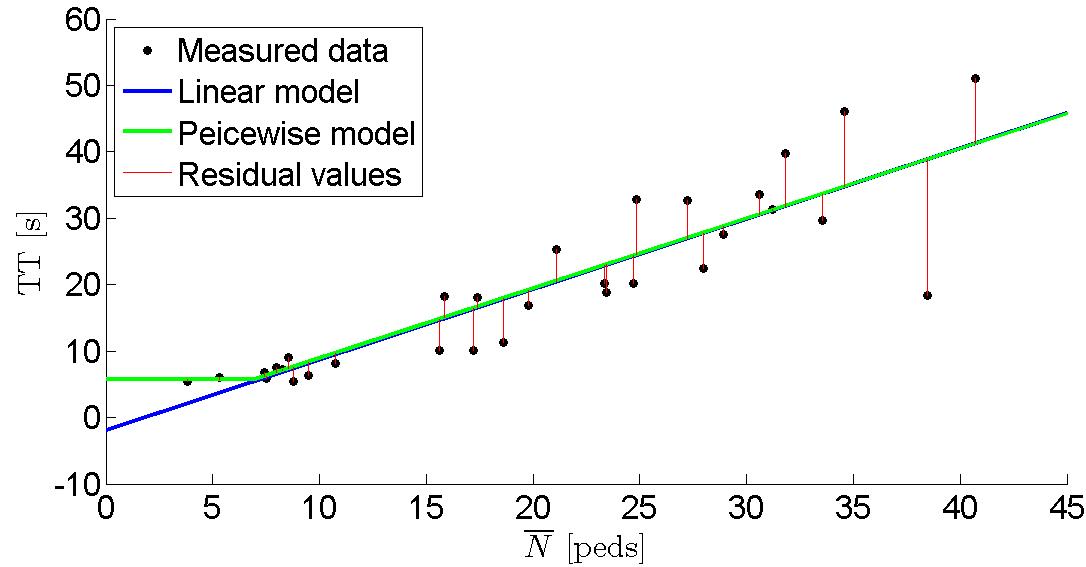}
		\phantom{x}
		\includegraphics[height=0.15\textheight]{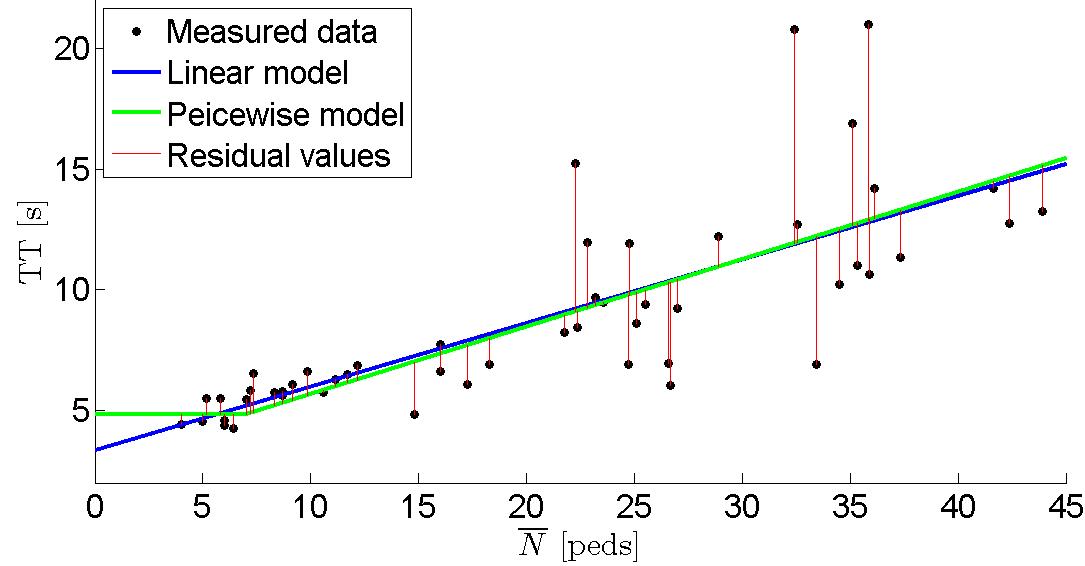}
	\end{center}
	\caption{Individual results of travel time linear model. For selected fast (right) and slow (left) participants, standard and piece-wise linear model are compared.}
	\label{fig:TT_model_ped}
\end{figure}

Comparing to classic linear model $TT(i_{\alpha}) = \frac{S}{v_0} + \overline{N}(i_{\alpha})$, piece-wise model fits the data much better mainly in free flow area, where the travel time obviously does not depend on occupancy. Vice versa, the area of constant TT trend does not affect the slope modeled in crowded area, see Figure \ref{fig:TT_model_ped}.

Mean value of residuals in both models are similar (see Table \ref{tab:TT_residua}), the lower value of mean $\mathrm{R}^2$ for piece-wise model corresponds to the facts mentioned above.

\begin{table}[h!]
	\caption{Table $\mathrm{R}^2$ values for piece-wise and standard linear models evaluated for each pedestrian.}
	\label{tab:TT_residua}
	\centering
	\begin{tabular}{p{1.2cm}p{1.2cm}p{1.2cm}p{1.2cm}|p{1.2cm}p{1.2cm}p{1.2cm}p{1.2cm}}
	\multicolumn{4}{c|}{Piece-wise model} & \multicolumn{4}{c}{Standard linear model}\\
	\hline
	mean  \centering &median \centering  &min \centering  &max \centering   &mean  \centering &median \centering  &min \centering   &max 	\\
	0.688 \centering &0.691 \centering   &0.386\centering &0.936\centering  &0.679\centering  &0.676\centering    &0.362\centering  &0.938	
	\end{tabular}
\end{table} 

\begin{figure}[h!]
	\begin{center}
		\includegraphics[height=0.13\textheight]{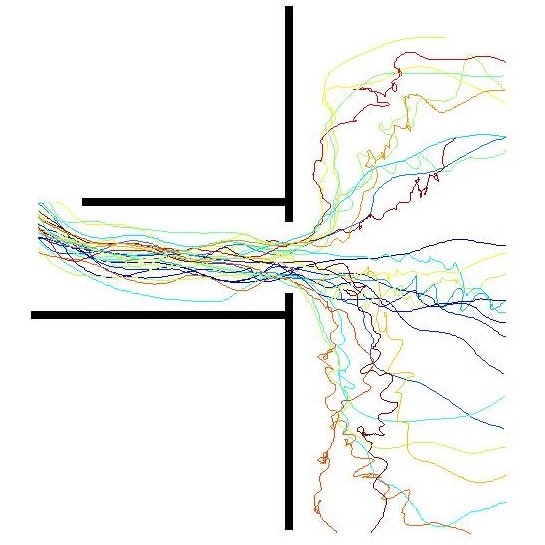} 
		\phantom{x}
		\includegraphics[height=0.13\textheight]{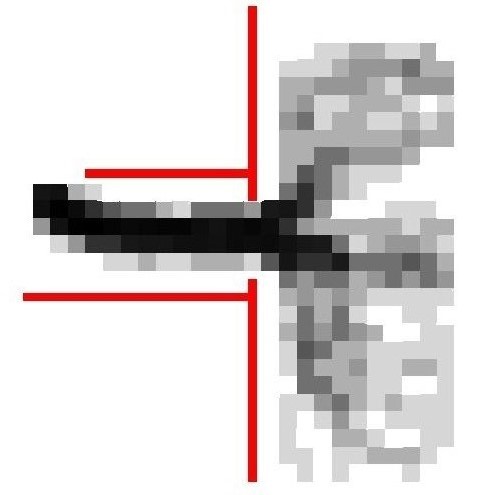} 
		\phantom{x}
		\includegraphics[height=0.13\textheight]{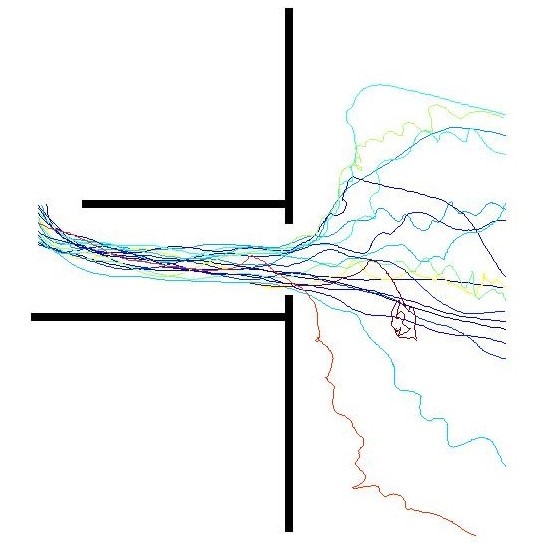}
		\phantom{x}
		\includegraphics[height=0.13\textheight]{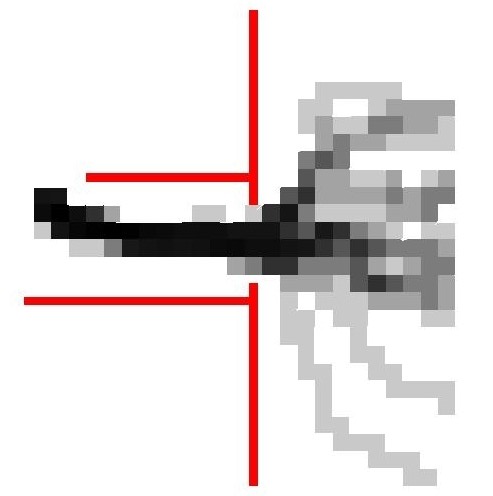} 
	\end{center}
	\caption{Paths and path density of slow pedestrian (right) and fast one (left).}
	\label{fig:Paths_individual}
\end{figure}

At the end, the path density was investigated with respect to individual behavior, see Figure \ref{fig:Paths_individual}. The observed trends are similar to the conclusions drafted from anonymous paths data.   

To conclude this part, observed variance is successfully explained by differences in individual behavior. Pedestrians hold to their different strategies which leads to different performances.

\section{Conclusions}

Even it was shown that lover travel time was reached by faster trajectories, the paths selection itself would not imply fast passing. Fast paths were not shorter or better curved to reach low travel time from physical point of view. But this paths were more effective to get through the dense crowd. 

To support this idea, we found out many tokens of aggressive behavior from the camera records as pushing, rude overtaking or blocking each other. The concept of aggressiveness as a feature of pedestrian fits this idea well. This is supported by the fact that some individuals reached low travel time under all conditions while others were very sensitive to occupancy. 

The effect of heterogeneity has dramatic influence to the progress of evacuation. The time spent in dangerous area may be in mean value sufficiently low according to local guidelines, but there is a high probability that some part of pedestrians would stay there for much longer. This effect was implemented to the cellular automata model, see \cite{PPAM14}.

\section*{acknowledgement}
This work was supported by the Czech Science Foundation under the grant GA15-15049S and by Czech Technical University under the grant SGS15/214/ OHK4/3T/14. Experimental records available at the link https://www.youtube.com/watch?= d4zZpvhahYM.

\end{document}